# Inverted Classroom an der Hochschule Karlsruhe - ein nicht quantisierter Flip

Isabel Braun, Gottfried Metzger, Stefan Ritter, Mikko Vasko, Hans-Peter Voss

Projekt SKATING, Hochschule Karlsruhe

Ab wann ist eine Veranstaltung 'invertiert'? Speziell beim Erstkontakt mit einer neuen Methode wird ein reflektierender Dozent - auch im Sinne seiner Studierenden - zunächst versuchen diese auszuprobieren, ohne die ihm bekannten Vorteile des bewährten Systems sofort aufzugeben. Anschließend muss er bewerten, ob das neue System Erfolg hatte und einen Mehrwert für Lernende und/oder den Lehrenden erzeugt. Er wird sich der Frage stellen müssen, wann eine invertierte Veranstaltung erfolgreich ist. Im Hinblick auf eine objektive Bewertung des Resultats und die tägliche Arbeitsbelastung der Lehrenden kann es für eine Umstellung nützlich sein, wenn diese von einer dritten Partei beobachtet und begleitet wird.

Im Rahmen des durch den ''Qualitätspakt Lehre'' (BMBF) finanzierten Projektes SKATING an der Hochschule Karlsruhe – Technik und Wirtschaft wird unter anderem individuelle technische und didaktische Beratung für Lehrende angeboten. Neben Unterstützung bei der Umsetzung von e-Learning-Angeboten beinhaltet dies auch konkrete Hilfen zur Entwicklung, Implementation und Erprobung innovativer Lehrmethoden wie z.B. dem Inverted Classroom Modell.

In Gesprächen mit Professoren wurde Interesse an dieser Methode deutlich. Dadurch wurden mehrere Pilotprojekte angeregt, in denen Varianten des Inverted Classroom umgesetzt und evaluiert werden sollen. Bei diesen Vorhaben wird großer Wert auf die Anpassung an die jeweiligen Bedürfnisse der verschiedenen Professoren, Kurse und Studierenden gelegt. Das Herantasten an die neue Methode wird anhand von unterschiedlichen konkreten Beispielen skizziert.

## 1.1 Einführung

Die *Hochschule Karlsruhe – Technik und Wirtschaft* ist im Jahr 2005 aus der Fachhochschule Karlsruhe – Hochschule für Technik hervorgegangen. Ursprünglich gegründet wurde die Lehranstalt bereits 1878 unter dem Namen „Großherzogliche Badische Baugewerkeschule".

An sechs Fakultäten werden ca. 7 000 Studierende in 37 Studiengängen der Ingenieurwissenschaften, der Wirtschaftswissenschaften und der Informatik ausgebildet, wobei 20,7% der Studierenden weiblich und 14,1% ausländischer Herkunft sind. Betreut werden sie von 183 hauptamtlichen Professorinnen und Professoren, 19 Honorarprofessoren sowie von 329 Lehrbeauftragten und über 400 Mitarbeitern (Stand April 2012). Kontinuierlich steigende Bewerberzahlen und hervorragende Ergebnisse in wichtigen nationalen Hochschulrankings belegen den akademischen Erfolg.

### 1.1.1 Probleme der Studierenden

Im Bereich der Ingenieurwissenschaften wird vor einem zunehmenden Fachkräftemangel gewarnt (z.B. [INT01]). Zu wenige junge Menschen entscheiden sich für ein ingenieurwissenschaftliches Studium. Vor allem weibliche Studieninteressierte wählen auch heute seltener einen solchen Studiengang als ihre männlichen Altersgenossen. Hinzu kommt, dass nicht alle Studierenden das begonnene Studium erfolgreich abschließen. Hohe Dropout-Quoten in ingenieurwissenschaftlichen Studiengängen (z.B. Heublein, et al., 2010) sind ein zentrales Problem, sowohl als individuelle Schicksale, wie auch volkswirtschaftlich.

Die Abbruchquoten auf der Makroebene sind das augenfälligste Indiz für ausbleibenden Studienerfolg, jedoch nur das letzte Glied einer langen Kette von Ursache-Wirkungs-Beziehungen. Ob die selbst initiierte vorzeitige Beendigung eines begonnenen Studiums oder eine automatische Exmatrikulation aufgrund von nichterbrachten Prüfungsleistungen, Abbrüche sind meist multifaktoriell bedingt.

Um den Abbrüchen bereits im Vorfeld mit Maßnahmen begegnen zu können, müssen die Ursachen bekannt sein. Aus diesem Grund werden am *Service-Center Studium und Lehre* an der Hochschule Karlsruhe verschiedene Anstrengungen unternommen, diese zu identifizieren. Insbesondere durch das im Rahmen des *Qualitätspakts Lehre* durch das Bundesministerium für Bildung und Forschung geförderte Projekt *SKATING* konnten diese Bemühungen um eine Aufklärung von Abbruchursachen weiter intensiviert werden. Es wurden und werden Gespräche mit Studierendenvertretern, Lehrenden und Schlüsselpersonen verschiedener Hochschuleinrichtungen sowie schriftliche Befragungen durchgeführt. Außerdem werden Dokumente und Datenbestände analysiert.

Insbesondere kommen Studierende oftmals mit der in Folge des Bologna-Prozesses erhöhten Anzahl an zu erbringenden Prüfungsleistungen zum Semesterende nicht optimal zurecht. Von Lehrenden werden als Hauptursachen hierfür ein Mangel an semesterbegleitendem Lernen und ein Defizit in der systematischen Vor- und Nachbereitung von Studienveranstaltungen genannt. Nach den Ursachen für den Verzug von Prüfungsleistungen befragt, räumen Studierende oftmals ein, den Lernaufwand unterschätzt zu haben und schreiben Misserfolge dann zu geringen eigenen Anstrengungen und fehlenden Lernstrategien zu. In Übereinstimmung mit der Einschätzung vieler Lehrender geben sie an, selten Lehrveranstaltungen vor- oder nachbereitet und erst spät mit dem Lernen für Prüfungen begonnen zu haben.

Eine zentrale Herausforderung mit der sich alle Hochschulen für Angewandte Wissenschaften in besonderem Maße konfrontiert sehen, ist darüber hinaus die hohe Heterogenität der Eingangsvoraussetzungen ihrer Studienanfänger, die sich aufgrund der weiter zunehmenden Vielfalt der Zugangswege ergibt. Diese Öffnung der Hochschulen soll mit dazu beitragen, dem Fachkräftemangel im Ingenieurbereich entgegen zu wirken. Was politisch gewollt ist, erweist sich in der Praxis der Hochschulen als problematisch: Es ist schwierig, dem unterschiedlichen Kenntnisstand der Studienanfänger und den daraus resultierenden Unterschieden in der Lerngeschwindigkeit gerecht zu werden.

Wie kann mit unterschiedlichen Lerngeschwindigkeiten umgegangen werden? Auf welchem Wege können Studierende beim semesterbegleitenden Lernen unterstützt und zur kontinuierlichen aktiven Wissensaneignung bereits während der Vorlesungszeit motiviert werden? Wie kann die Hochschullehre unter den gegebe-nen Rahmenbedingungen so gestaltet werden, dass durch sie den Studierenden ein möglichst hoher Mehrwert erwächst?

### 1.1.2 Das Projekt 'SKATING'

Durch die hohe zeitliche Auslastung der Lehrenden und den weitgehend fehlenden Mittelbau an den Hochschulen für Angewandte Wissenschaften bleiben im regulären Veranstaltungsbetrieb kaum die Ressourcen, sich mit Details der Umsetzung möglicher Lösungsansätze für die genannten Probleme zu befassen. Die vom Bundesministerium für Bildung und Forschung im Rahmen des Qualitätspakts Lehre deutschlandweit geförderten Projekte an Hochschulen eröffnen vielerorts Freiräume, um sich mit diesen Schwierigkeiten auseinanderzusetzen und gegebenenfalls innovative Lösungsansätze zu planen, umzusetzen und zu evaluieren.

Unter der Bezeichnung „SKATING" (Studienreformprozess Karlsruhe zur Transformation des Ingenieurstudiums) wird an der Hochschule Karlsruhe in enger Zusammenarbeit mit der Geschäftsstelle der Studienkommission für Hochschuldidaktik (GHD) ein solches Projekt umgesetzt. Ziel ist es, die Studienbedingungen in den ingenieurwissenschaftlichen Studiengängen weiter zu verbessern und insbesondere die Qualität der Hochschullehre so weiterzuentwickeln, dass sie den aktuellen Herausforderungen an den Hochschulen für Angewandte Wissenschaften möglichst gerecht wird.

Basierend auf anfänglichen Bestandsaufnahmen und Bedarfsanalysen werden im Rahmen von Pilotprojekten gemeinsam mit den Lehrenden innovative Lehrmethoden erprobt. Durch die enge Zusammenarbeit mit den Lehrenden entstehen auf die Bedürfnisse und Persönlichkeit des Dozenten abgestimmte Unterrichtsmodelle, die stets lebendig weiterentwickelt oder auch wieder verworfen werden können, denn letztlich entscheidet die Zufriedenheit des Lehrenden mit der eingesetzten Methode über Art und Dauer der Umsetzung

### 1.1.3 Lehr- oder Lernveranstaltung? - Die Inverted Classroom Methode

**Zur Bedeutung der Lehrform ‚Vorlesung'**

Im Zuge der allgemeinen Anerkennung der Notwendigkeit eines „Shift from Teaching to Learning" in der Hochschullehre wird heute gern die traditionelle Frontalvorlesung als hochschuldidaktischer Anachronismus aufgefasst, der schleunigst durch „Aktivierende Veranstaltungsformen" zu ersetzen sei. Hierbei ist jedoch schon die Unterstellung, eine Vorlesung könne per se keine „Aktivierende Veranstaltungsform" sein, in dieser Pauschalierung falsch. Mit Aktivierung ist nicht einfach ein äußerlicher Aktivismus gemeint, sondern auch und gerade das geistige Tätigsein des Lernenden. Es gibt Lehrende, denen eine solche innere Aktivierung ihres Publikums selbst über längere Zeiträume hinweg hervorragend gelingt. Allerdings mag auch ihnen diese Aktivierung noch besser gelingen, wenn im Unterricht rezeptive und eigenproduktive Phasen der Studierenden einander ergänzen und rhythmisch abwechseln.

Welche Rolle spielt die Frontalvorlesung in einem typischen mathematisch-naturwissenschaftlichen oder ingenieurwissenschaftlichen Studium? Sie ist besonders an Universitäten gerade keine „Lernveranstaltung" im eigentlichen Sinne, sondern eine Stoffpräsentationsveranstaltung, in welcher der „State of the Art" der Disziplin auf in der Regel hohem Abstraktionsniveau vorgestellt wird. Sie gibt einen Ausblick auf den mehr oder weniger kanonisierten Stand des Faches. Sie gibt Überblicke, entwickelt den Roten Faden des Stoffes und ist die fachliche Referenz.

Eigentlich gelernt wird der Stoff erst durch ein intensives Selbststudium, das neben der Vorlesung noch andere Quellen hinzuzieht, durch die selbständige Bearbeitung von Fragen und Übungen zum Stoff, durch deren Besprechung in Übungsgruppen, durch Praktikums- und Laborveranstaltungen und schließlich durch selbst organisierte Lernpartnerschaften.

Erst die Gesamtheit dieser Lernkontexte sorgt dafür, dass Kenntnisse nachhaltig angeeignet werden, Verständnis solide aufgebaut wird und komplexes Methodenwissen auch im Transfer beherrscht wird.

**‚Inverted Classroom' als Lernveranstaltung**

Inverted Classroom (IC) begreift sich als eine Umkehrung der oben dargestellten traditionellen Lehrstruktur. Als ein Wesensmerkmal dieser klassischen Struktur wird der Umstand betrachtet, dass zunächst die Wissensvermittlung in der Vorlesung stattfindet, danach dann die Aneignung des Wissens in Eigenarbeit oder auch in Lerngruppen.

Die größten Schwierigkeiten treten häufig nicht beim Erstkontakt mit dem Stoff bei seiner Vermittlung auf, sondern beim tieferen Erfassen und Begreifen der Inhalte und Konzepte sowie bei der Anwendung auf praktische Probleme, und damit insbesondere dann, wenn der Lernende auf sich gestellt ist.

Möchte man Studierende in der Aneignungsphase unterstützen, ist es unter der Annahme, dass die Reihenfolge: Wissensvermittlung – gründlichere Aneignung - nicht verändert werden kann, ein naheliegender Schritt, die erste Phase (die Wissensvermittlung) im Rahmen eines vorgelagerten Selbststudiums zu verlangen.

Zentrales Merkmal der ICM ist somit, dass - im Gegensatz zu traditionellen Vorlesungen - die aktive Wissensaneignung, der eigentliche Lernprozess also, innerhalb der Präsenzveranstaltungen stattfindet. Die an passive Subjekte gerichtete lehrendenzentrierte Wissensvermittlung wird aus der Präsenzveranstaltung ausgelagert, in der damit Raum geschaffen wird für die praxisnahe Anwendung neu erworbener Wissensinhalte, zum Üben, zum Vertiefen, für aktive Problemlöseprozesse und vor allem für den echten Austausch mit dem Lehrenden.

Auf der Suche nach Lehrmethoden, die der oben geschilderten Ausgangslage unter den Studierenden gerecht werden, fiel das Augenmerk daher auch auf die Inverted Classroom Methode (ICM).

**Die zwei Seiten des ICM**

Bei der Umsetzung des ICM stellen sich zwei wesentliche Fragen: „Wie realisiere ich die Wissensvermittlung?" und „Wie erzeuge ich einen deutlichen Mehrwert für die Präsenzveranstaltung, ohne neue Inhalte zu präsentieren?"

Eine sehr gute Inspiration für die Nutzung der Präsenzphase zur Förderung eines tieferen Verständnisses am Beispiel der Physik findet sich im Buch "Peer Instruction" von Eric Mazur (Mazur, 1997). Dieses in Harvard entwickelte Konzept beinhaltet ebenfalls bereits die didaktische Methode, die Informationsvermittlung aus der Vorlesung auszulagern, und motivierte viele didaktisch interessierte Dozenten der Physik zu konzeptorientiertem Unterricht mit erweiterter Interaktivität (Abstimmungen und Diskussionen über Multiple-Choice Verständnisfragen mit Clickern bzw. Flashcards). Da im Gegensatz zu heute schnelles Internet noch nicht umfänglich verfügbar war und der Schwerpunkt des Konzeptes auf der Umgestaltung der Präsenzphase lag, bekamen die Studierenden vor der Lehrveranstaltung einen Leseauftrag, dessen Erfüllung durch einen kurzen Test überprüft wurde.

Durch den Zuwachs an technischen Möglichkeiten und die immense Vielfalt an online verfügbaren Informationen erweitert sich aktuell das Potential der Methode. Die Bereitstellung von Videomaterialien versetzt Studierende in die Lage, die Geschwindigkeit der Wissensvermittlung selbst zu steuern. Auch wenn auf Rückfragen oder Verständnisprobleme nicht mehr direkt durch den Lehrenden eingegangen werden kann, hat der Zuhörer nun die Möglichkeit, die Aufzeichnung für eigene Recherchen zu unterbrechen. Er kann so kritische Passagen wiederholen oder verbleibende Fragen - beispielsweise in einem Diskussionsforum - an den Lehrenden zur Vorbereitung der nächsten Präsenzveranstaltung übermitteln. Fast alle Aspekte, die eine Vorlesung erfolgreicher machen als das Selbststudium aus einem Buch, können mit einem Video simuliert werden.

Durch die zunehmende Verbreitung von Internetvideoportalen, wie YouTube, ist die Bereitstellung von Lernvideos und Vorlesungsaufnahmen auch in Hochschulen populärer geworden (siehe z.B. [INT02]).

**Beispiele aus den Lehrkulturen verschiedener Fächer**

Im Hochschulbereich unterscheiden sich die Lehrkulturen verschiedener Fachrichtungen in erheblichem Maße. So sind viele aktivierende Methoden wie „Blitzlicht" und „Partnerinterview" in der Vermittlung von Fremdsprachen eine Selbstverständ-lichkeit, während sie im Bereich der Vermittlung mathematisch-naturwissenschaftlicher oder technischer Inhalte eine didaktische Innovation darstellen können. Analoges gilt auch für die ICM. Sie kann in diesen Feldern neue Perspektiven für das Lehren und Lernen eröffnen.

Historisch gesehen ist das geschilderte Umkehrkonzept vielerorts eine akademische Selbstverständlichkeit, etwa in Seminaren zur Philosophie oder in den Literaturwissenschaften, wo das Lesen eines behandelten Textes in Eigenverantwortung vor der Unterrichtsstunde erfolgt, die dann der Besprechung von Fragen, der Interpretation und der Diskussion gewidmet ist (vgl. Handke, 2012).

Toto & Nguyen (2009) sehen die ICM als Möglichkeit, trotz einer großen Stofffülle, die nicht ohne eine frontale - hier multimedial unterstützte - Vermittlung auskomme, in Präsenzveranstaltungen eine aktive selbstgesteuerte, bestenfalls kooperative Auseinandersetzung mit Lerninhalten zu ermöglichen. Sie betonen den Vorteil, dass mit der Methode in den Präsenzveranstaltungen mehr Raum für direkte Interaktion sowohl unter den Studierenden als auch zwischen den Studierenden und der Lehrperson möglich wird. Die Studierenden haben die Möglichkeit, Fragen, die sich während der Wissensvermittlung aufgetan haben, während der Wissensaneignung zu stellen. Lehrende können direkt Feedback geben oder die Leistung und die Lernfortschritte direkter beobachten, wie dies beispielsweise Gannod et al. (2008) darstellen.

**Zusätzlicher Nutzen: Genderaspekt**

Speziell für ingenieurwissenschaftliche Studiengänge könnte die ICM auch deshalb ein interessanter Ansatz sein, weil er geeignet sein könnte, die Attraktivität der Veranstaltungen für weibliche Studieninteressierte zu steigern, ohne dadurch männliche abzuschrecken. Hinweise auf Geschlechtsunterschiede in der Akzeptanz von ICM finden sich beispielsweise bei Lage et al. (2000). Die ICM wurde in dieser Studie über alle Studierenden hinweg positiv bewertet, die Bewertung durch weibliche Studierende fiel jedoch positiver aus, als die der männlichen Kommilitonen. Lage et al. (2000) stellen die Hypothese auf, dass insbesondere der Aspekt der kooperativen Lernformen in den Präsenzveranstaltungen beim ICM bei weiblichen

Studierenden Anklang findet. Zu einem ähnlichen Schluss kommen auch Lorenzo et al. (2006), die den positiven Einfluss des verwandten Peer Instruction Konzeptes auf das inhaltliche Verständnis geschlechtsspezifisch untersuchten. Durch interaktive Lehrmethoden konnte ein anfangs bestehender Unterschied im Stoffverständnis reduziert oder sogar eliminiert werden, ein Effekt, welcher der kommunikativen Ausrichtung dieser Methode zugeschrieben wird.

**Umsetzung der ICM**

Der Einstieg in die ICM mit Videoeinsatz ist zunächst mit einem gewissen Arbeitsaufwand verbunden. Nach dem erstmaligen Erstellen der Materialien und der Institutionalisierung der notwendigen Prozesse und Arbeitsschritte ist in den Folgesemestern jedoch mit einem deutlich verringerten Aufwand zu rechnen. Die finanziellen Kosten und der Arbeitsaufwand für die Erstellung und Vervielfältigung qualitativ hochwertiger audiovisueller Medien sind überdies in den letzten Jahren stetig zurückgegangen. Insbesondere kann sich in Antizipation eines möglichen zukünftigen Einstiegs in die ICM die Aufzeichnung der „traditionellen" Vorlesungen in den vorausgehenden Semestern lohnen, wie dies beispielsweise Lage et al. (2000) empfehlen.

## 1.2 Evaluation der ICM

Keine innovative Lehrmethode sollte um ihrer selbst Willen eingesetzt werden oder weil man „schon immer mal eine bestimmte Technologie ausprobieren wollte". Ausgangspunkt sollte immer ein im Voraus zu bestimmender Zielzustand, in Gegenüberstellung zu einem momentanen Ist-Zustand, sein. Zur Erreichung des erwünschten Zielzustandes werden unter Beachtung der jeweiligen Randbedingungen geeignete Maßnahmen, hier Lehrmethoden, ausgewählt. Um die voraussichtliche Eignung einer Methode vorab zu beurteilen, werden auf Basis einer Theorie oder eines Modells Hypothesen über mögliche Wirkungszusammenhänge aufgestellt, bestenfalls durch empirische Ergebnisse unterstützt.

Mit der Einführung jeder neuen Lehrmethode sind somit bestimmte Erwartungen, das Eintreten eines mehr oder minder genau spezifizierten Mehrwertes verbunden. Ausgehend von der oben geschilderten Problemlage wurde im vorliegenden Fall die ICM als eine potenziell geeignete Methode identifiziert.

Dass mit einer Maßnahme aber auch tatsächlich das Intendierte erreicht wird, sollte auch im Falle einer fundierten Maßnahmenauswahl nicht von vornherein als gegeben betrachtet werden, sondern Gegenstand empirischer Untersuchungen sein. Ziel von *Evaluation* ist es zum einen *summativ* zu bewerten, inwieweit die Erwartungen eingetreten sind und abschließend einzuschätzen, ob der Einsatz der neuen Methode erfolgreich war. Darüber hinaus sollte aber auch prozessbegleitend *formativ* evaluiert werden, um im Bedarfsfall nachjustieren und eventuellen unintendierten negativen Effekten begegnen und ein Programm gegebenenfalls abbrechen zu können.

*Evaluationsforschung* im engeren Sinne bezeichnet den Einsatz wissenschaftlicher Forschungsmethoden und Modelle um eine solche Bewertung vorzunehmen (vgl. Wittmann, 1985). Letztendlich geht es darum, durch Evaluation Entscheidungen zu ermöglichen, in diesem Fall z.B. inwieweit eine Lehrmethode als erfolgreich weiter eingesetzt oder inwiefern Modifikationen vorgenommen werden sollten. Unter dieser Perspektive beinhaltet Evaluation alle Aktivitäten der „Sammlung, Analyse, Interpretation und Kommunikation der Arbeitsweise und Effektivität von Programmen, Produkten und Interventionen" (Wittmann, 2009, S. 60).

Evaluation wird somit zu einem unverzichtbaren Bestandteil von Interventionsvorhaben. Die jeweilige Evaluationsstrategie, die Wahl der Methoden und Kriterien, ist auf das jeweilige Programm maßzuschneidern, um dessen Erfolg angemessen beurteilen zu können.

**Wahl der Bewertungskriterien**

Eine entscheidende Frage ist, aus welcher Perspektive der Erfolg bewertet wird. In der Regel sind die nicht immer miteinander zu vereinbarenden Interessen und Erwartungen unterschiedlicher Anspruchsgruppen gleichzeitig zu berücksichtigen. Sich diese zu vergegenwärtigen sollte der Ausgangspunkt jedes Evaluationsvorha-bens sein. Im vorliegenden Fall sind die Lehrenden, die Studierenden, die Hochschule als Institution, der Mittelgeber oder auch die Wissenschaftler des Projekts zu nennen. Beispielsweise sind zur Bewertung der ICM aus Sicht des Lehrenden dessen Vorstellungen von einer gelungenen Lehrveranstaltung und damit die individuellen Kriterien für den Erfolg auf organisatorischer, fachlicher, methodischer und persönlicher Ebene zu erheben.

Aus der Stakeholder-Perspektive lassen sich Zielkriterien und erste Überlegungen zu deren Operationalisierung ableiten, um die Wirkung und den Erfolg einer Maßnahme bewerten zu können. Zur Bewertung des Erfolgs der ICM könnte als Ergebniskriterium unterschiedlicher Anspruchsgruppen letztendlich „Studienerfolg" der teilnehmenden Studierenden benannt werden, ob nun verstanden als „die Beendigung des Studiums mit bestandener Abschlussprüfung", „gute Studiennoten", „wenig Prüfungswiederholungen", „eine kurze Studiendauer", „eine hohe Studienzufriedenheit", „der Erwerb berufsrelevanter Kompetenzen" oder auch „anschließender Berufserfolg" (vgl. z.B. Rindermann & Oubaid, 1999, S. 175, Hell et al., 2008, S. 44). Zweifellos nachvollziehbar ist die Empfehlung, anstelle solch globaler und distaler Kriterien den Erfolg einer Einzelintervention, wie der Einsatz der ICM im Rahmen eines Kurses, anhand von proximalen, dem Kursgeschehen nahen Ergebniskriterien zu bewerten. In Frage kommen beispielsweise der Kompetenzgewinn in den behandelten Themenbereichen oder das Erreichen von vorab für die Lehreinheit definierten Lernzielen.

Die Wahl sollte sich auch nicht auf ein singuläres Ergebniskriterium beschränken, sondern breit gefächert Kriterien ausgewählt werden, die möglichst alle Aspekte einer Intervention angemessen widerzuspiegeln (vgl. Wittmann, 2009). Beispielsweise können im vorliegenden Fall das Ebenen-Modell von Kirkpatrick (1979) oder etablierte Lernzieltaxonomien eine Planungshilfe darstellen.

**Fragen der Datenerhebung**

Noch nicht geklärt ist damit die Frage, mit welchen Methoden oder über welche Datenquellen die Ausprägung auf Zielvariablen erhoben wird. Als Datenquellen kommen Fragebogendaten, beispielsweise Selbstauskünfte der Studierenden (z.B. bei Lage et al. 2000; Gannod et al., 2009), Ergebnisse objektiver Tests, beispielsweise standardisierte Konzepttests (z.B. bei Papadopoulos et al., 2010) oder maßge-schneiderte kriterienorientierte Tests, die Einschätzung des Lehrenden oder eines anderen Beobachters in Frage.

Bestenfalls werden nicht nur mehrere Kriterien, sondern auch unterschiedliche Datenquellen und Datenerhebungsmethoden kombiniert eingesetzt.

Um gefundene Effekte kausal auf eine Intervention zurückführen zu können, sind (quasi-) experimentelle Versuchspläne zu empfehlen. Im Idealfall würde dies zum einen die Erfassung der Ausgangsleistung vor der ICM-Lehreinheit, zum anderen den Einbezug einer Vergleichsgruppe implizieren. Damit auf deskriptiver Ebene gefundene Effekte, ob im Gruppenvergleich oder in Gegenüberstellung von Prä-Post-Messungen, angemessen auf statistische Signifikanz überprüft werden können, sind wiederum relativ große Untersuchungsgruppen notwendig. Ausschlaggebend für die Wahl des Untersuchungsdesigns sind letztendlich oftmals pragmatische Überlegungen bezüglich der verfügbaren Ressourcen.

Zusätzlich ist der Zeithorizont zu beachten: was sind kurz-, mittel- und langfristige Effekte? Gegebenenfalls werden im Rahmen von ICM vermittelte Inhalte besonders tief verarbeitet, so dass sie insbesondere auch nach längeren Zeiträumen abgerufen werden können. Die Überlegenheit von ICM gegenüber traditionellen Lehrmethoden würde sich dann insbesondere in wiederholten Follow-up-Messungen nach längeren Zeiträumen zeigen.

**Unintendierte Effekte**

Die Wirkung von ICM umfassend zu untersuchen, impliziert auch, den Blick nicht auf die Betrachtung der erwünschten Effekte, deren Eintreten oder Ausbleiben, zu verengen, sondern mögliche Risiken und negativen Folgen zu beachten. Unintendierte Effekte von ICM könnten sich beispielsweise durch den hohen Zeitbedarf auf Seiten der Studierenden ergeben. Gegebenenfalls leiden Studienleistungen in anderen Bereichen unter der erhöhten zeitlichen Belastung durch die ICM-Kurse. So berichten Papadopoulos et al. (2010), dass die Studierenden nach eigener Angabe mehr Zeit für den ICM-Kurs aufwenden mussten als für andere Kurse mit vergleichbarer Anzahl an Credit-Punkten. Auch auf Seiten der Lehrenden und anderen an der Umsetzung Beteiligten sollten der Aufwand und die Arbeitsbelastung durch diese Methode betrachtet und gegebenenfalls Kosten-Nutzen-Analysen durchgeführt werden. Ob sich für den Lehrenden, wie oben vermutet, zumindest langfristig eine Arbeitsentlastung gegenüber der klassischen Veranstaltungsgestaltung ergibt, bleibt zu prüfen, denn auch für die interaktiven Präsenzveranstaltungen erwarten Studierende eine gute Vorbereitung und Organisation durch den Lehrenden (vgl Toto & Nguyen, 2009). Beispielsweise berichten Haden et al. (2009), dass für die Erstellung von 20 Minuten Videomaterial für die Nutzung im Rahmen von ICM mindestens drei Stunden Zeit investiert werden mussten.

Den Blick nicht nur auf die erwarteten und erhofften Effekte zu verengen, lässt sich insbesondere dann umsetzen, wenn neben standardisierten Tests und geschlossenen Fragen zusätzlich Methoden aus dem Bereich der qualitativen Forschung eingesetzt werden (vgl. bei Haden et al., 2009), beispielsweise unstrukturierte Gespräche mit Beteiligten durchgeführt oder Informationen durch teilnehmende Beobachtung erhoben werden. Insbesondere wenn zu einer Interventionsmethode bisher wenige Erfahrungen vorliegen, sollte nicht auf solche hypothesengenerierenden Verfahren verzichtet werden. Zumindest sollte jedoch in Fragebogenverfahren Raum für offene Anmerkungen und Verbesserungsvorschläge eingeräumt werden.

**Einfluss weiterer Variablen**

Darüber hinaus darf nicht ungeprüft davon ausgegangen werden, dass alle Beteiligten in gleichem Maße und unabhängig von den jeweiligen individuellen Voraussetzungen und spezifischen Randbedingungen von einer neuen Lehrmethode profitieren. Dies ist die Frage nach differentiellen Effekten und der Wirkung von Moderatoren. Als Randbedingungen könnte beispielsweise die Art der Hochschule, das Studienfach, das Thema des Kurses oder die Größe der Gruppe eine Rolle spielen. Erfolgreich eingesetzt wurde die ICM sowohl in den Wirtschafts- (vgl. z.B. Lage et al., 2000) wie auch den Ingenieurwissenschaften (vgl. z.B. Papadopoulos et al., 2010, Toto & Nguyen, 2009).

Auf Seiten der Studierenden sind insbesondere in Anbetracht der hohen Heterogenität der Studierendenschaft differentielle Effekte im vornherein keinesfalls auszuschließen. Unterschiede in Abhängigkeit von dem Geschlecht der Studierenden wurden bereits oben dargestellt. Darüber hinaus könnten motivierte Studierende mit Defiziten im mitgebrachten Schulwissen in besonderem Maße von der Möglichkeit zur Wissensvermittlung in eigener Lerngeschwindigkeit profitieren. Neben soziodemographischen Variablen und Unterschieden im Wissensstand kommen auch Persönlichkeitsunterschiede als erklärende Variable in Frage. Unabhängig davon ob man sich auf die Diskussion um die Sinnhaftigkeit oder den heuristischen

Nutzen von Lernertypologien einlassen möchte, ist damit zu rechnen, dass Studierende mit unterschiedlichen Ausprägungen auf verschiedenen Persönlichkeitsdimensionen in unterschiedlichem Maße auf verschiedene Lehr- und Lernumwelten reagieren. Beispielsweise finden Toto & Nguyen (2009) Hinweise auf Zusammenhänge zwischen der individuellen Ausprägung auf verschiedenen Dimensionen eines Lernstil-Fragebogens, dem Nutzungsverhalten und der Bewertung der ICM. Vielleicht bietet diese Lehrmethode aber dennoch den Vorteil, dass im Vergleich zu traditionellen Lehrmethoden letztendlich alle Lernertypen profitieren, wie beispielsweise Lage et al. (2000) annehmen. Zusätzlich kann auch die Person des Lehrenden eine Rolle spielen, insbesondere in Bezug auf seine Begeisterung für die Methode und auf die Frage, inwieweit die ICM als neue Lehrmethode mit seinen bisherigen Erfahrungen, seiner allgemeinen Lehrphilosophie, seinen fachdidaktischen Vermittlungskonzepten und seinem Rollenverständnis kompatibel ist.

Letztendlich sind Einzelstudien jedoch nur begrenzt in der Lage solche Moderatoreffekte aufzudecken. Hierfür sind Metaanalysen notwendig. In Einzelstudien können nur die Voraussetzungen dafür geschaffen werden, dass eine spätere Integration der empirischen Einzelbefunde möglich wird. Dies impliziert nicht nur ein forschungsmethodisch einwandfreies Vorgehen, sondern insbesondere auch dessen detaillierte Dokumentation und die umfassende Beschreibung aller Randbedingungen und untersuchten Personengruppen.

**Prozessgeschehen**

Zur umfassenden Evaluation ist darüber hinaus auch unverzichtbar das eigentliche Interventionsgeschehen zu betrachten und die Blackbox zwischen Eingangsvoraussetzungen und Ergebniskriterien aufzubrechen. Auf diesem Wege können die Ursachen für das Ausbleiben von erwünschten Effekten oder das Auftreten von unerwünschten Wirkungen aufgedeckt werden. Im Sinne der Implementationskontrolle ist zu prüfen, inwieweit eine Lehrmethode wie intendiert umgesetzt wurde. Gegebenenfalls sind prozessbegleitend Maßnahmen zu ergreifen dies sicherzustellen. Beispielsweise sollte im vorliegenden Fall geprüft werden, inwieweit die Videos zur Vorbereitung rechtzeitig und in angemessener Qualität verfügbar waren oder ob technische Probleme auftraten. Studierende können zu Lehrmaterialien, zur Organisation der Durchführung, dazu, inwieweit ihnen die Aufgabenstellung klar war oder zur Passung der behandelten Inhalte für die ICM befragt werden (vgl. Lage et al., 2000; Haden et al., 2009).

Zusätzlich können weitere Informationen zum Prozessgeschehen erhoben werden, zum Ausmaß an Interaktion in den Präsenzveranstaltungen, zum Einsatz einzelner Lehrtechniken, zum Umstand, inwieweit auf Fragen eingegangen wird sowie zu den Fragen, ob Leistungsrückmeldung erfolgt, ob unterschiedliche Lerngeschwindigkeiten möglich sind und inwieweit durch die ICM semesterbegleitend gelernt wird. Diese Informationen können durch Befragung des Lehrenden, der Studierenden oder im Rahmen der Beobachtung durch Dritte erhoben werden.

Notwendige Bedingung für den Erfolg der ICM ist die Kooperationsbereitschaft der Studierenden. Diese müssen die Lehrmaterialien zur Wissensvermittlung vor den Präsenzveranstaltungen durcharbeiten, Videos anschauen oder Texte lesen und in die Veranstaltung kommen, worauf beispielsweise Gannod et al. (2009) hinweisen. Bestenfalls wird auf individueller Ebene das Ausmaß erfasst, in dem die Intervention angenommen wurde. Über Paradaten aus elektronischen Systemen oder über zeitnahe Befragungen lassen sich Informationen zum Nutzungsverhalten erheben, beispielsweise inwieweit die Videos konzentriert und vollständig angeschaut wurden, ob Ablenkungen auftraten, Passagen wiederholt oder wie viel Zeit insgesamt aufgewandt wurde (z.B. bei Toto & Nguyen, 2009). Diese Informationen können im Sinne einer „Interventionsdosis" interpretiert und für korrelative Auswertungen herangezogen werden. Die Akzeptanz der neuen Methode auf Seite der Studierenden ist von entscheidender Bedeutung. Daher sollte deren Zufriedenheit erfasst werden. In vielen Studien zur ICM werden die Studierenden beispielsweise am Ende des Semesters dazu befragt, welche Unterrichtsform sie bevorzugen. Zumeist schneidet die ICM in dieser Gegenüberstellung gut ab (z.B. Lage et al., 2000; Papadopoulos et al., 2010). Ähnliches gilt für die Bewertung der ICM durch die Lehrenden. Maßnahmen zu identifizieren, die Akzeptanz sicherstellen, ist von hoher Wichtigkeit. Der Einbezug der Studierenden in Planung und Bewertung der Methode erscheint aus dieser Perspektive unverzichtbar. Einen heuristischen Rahmen für Akzeptanzstudien zur ICM könnte das Technology Acceptance Model von Davis (z.B. 1989) oder eine seiner zahlreichen Modifikationen darstellen. Die detaillierte Beschreibung allen Prozessgeschehens ermöglicht zudem, zu untersuchen, auf welche Programmkomponente, welchen Aspekt, der Erfolg der ICM zurückzuführen ist. Möglich wäre beispielsweise, dass letztendlich die zusätzlich aufgewandte Lernzeit ausschlaggebend ist, oder es könnten die kooperativen Lernformen in den Präsenzveranstaltungen sein. Die Ausgestaltungsmöglichkeiten für die ICM sind in der Praxis vielfältig. Entscheidend ist, zu identifizieren, welche Komponenten für den Er-folg der Methode unter welchen Rahmenbedingungen unverzichtbar sind, um so die Planung und Umsetzung in der Praxis erfolgreich zu gestalten.

Für die Planung der eigenen Evaluation lassen sich die folgenden Punkte zusammenfassen:

- Prozessbegleitende Datenerhebung und zusammenfassende Bewertung
- Berücksichtigung der Interessen und Perspektiven aller Beteiligten
- Ergebniskriterien festlegen
    – Abgeleitet aus Stakeholderperspektiven
    – Passend zum Kursgeschehen
    – Multiple Ergebniskriterien
    – Unterschiedliche Datenquellen
    – Zu mehreren Zeitpunkten
- Mögliche unintendierte Effekte erfassen
    – Zeitbedarf auf Seite der Studierenden und Folgen daraus
    – Ressourcenbedarf für Planung, Vorbereitung, Implementation und Evaluation von ICM
    – Kombination von quantitativen & qualitativen Methoden; Verbesserungsvorschläge
- Einfluss weiterer Variablen untersuchen
    – Rahmenbedingungen
    – Untersuchungsstichprobe (Soziodemographie, kognitive Voraussetzungen, Persönlichkeitsvariablen)
    – Lehrperson
- Prozessgeschehen betrachten
    – Lehrmaterialien, technische Probleme, Organisation der Durchführung
    – Lehr- und Lernverhalten (z.B.: Interaktion zwischen Lehrendem und Lernenden, Beantworten von Fragen, kooperative Lernformen, semesterbegleitendes Lernen)
    – Nutzungsverhalten (Videos, Downloads, Einträge in Diskussionsforum o.Ä.)
    – Zufriedenheit, Vor- und Nachteile aus Sicht aller Beteiligten
    – Unterschiedliche Datenquellen (Befragung und Beobachtung Studierender und Lehrender )

## 1.3    Geplante ICM Varianten

Bei einem Physiker erzeugt das Wort Flip zumeist das Bild von wenigen diskreten Zuständen (ähnlich dem Spin eines Fermions in einem externen Magnetfeld), die einander stets ausschließen. Im Folgenden werden verschiedene Wege in eine ''invertierte'' Lehrveranstaltung aufgezeigt, die auch unterschiedliche Geschwindigkeiten in der Reformbereitschaft der Lehrenden zulassen, womit sich ganze Kurse sozusagen in Zwischenstadien befinden können.

Grundlage für die Bewertung des Systems ist die persönliche Vorstellung von einer gelungenen Lehrveranstaltung. Daher werden in Vorgesprächen die individuellen Kriterien für deren Erfolg auf organisatorischer, fachlicher, methodischer und persönlicher Ebene erhoben.

### 1.3.1    „Minimal-Invertiert": Aufzeichnung von Vorlesungszusammenfassungen

Diese Variante basiert auf Erfahrungen mit aufgezeichneten Vorlesungszusammenfassungen an der ETH Zürich (Schiltz 2011). Sie ist noch sehr nahe an der traditionellen Vorlesung, nur die Rekapitulationsphase, welche typischerweise am Anfang der Vorlesung stattfand, ist in den digitalen Raum verlagert und wurde als Podcast angeboten.

Pädagogische Vorteile dieser Form der Aufzeichnung sind:

1. mehr Zeit in der Vorlesung zur Vertiefung des Stoffes.
2. erleichterte Nachbereitung, insbesondere in der Prüfungsphase.
3. Verdeutlichung des 'Roten Fadens'.

Die Methode wie sie in Zürich eingesetzt wurde bedeutet jedoch einen Mehraufwand für den Dozenten, typischerweise in der Größenordnung einer Stunde pro Zusammenfassung (wöchentlich nach 4 Semesterwochenstunden) (Schiltz 2011). Da Professoren an Hochschulen für Angewandte Wissenschaften jedoch durch das hohe Lehrdeputat häufig nicht die Möglichkeit für einen solchen Mehraufwand sehen oder das persönliche Präsentieren der Zusammenfassung sehr schätzen, wurde die Methode für den Einsatz an der Hochschule Karlsruhe, der im Folgenden kurz beschrieben wird, entsprechend modifiziert. Die Zusammenfassung wird in der laufenden Vorlesung aufgenommen, womit noch immer zwei von drei Vorteilen der Züricher Form bewahrt werden.

Die Durchführung in einer Mathematikvorlesung begann im Sommersemester 2012. Der Dozent hält die ca. 10-minütigen Zusammenfassungen themenbezogen (in etwa) alle zwei Wochen. Aufgezeichnet und bearbeitet werden sie von einem Mitarbeiter des SKATING Projekts. Zu achten ist speziell auf gute Tonqualität, da die Aufzeichnung in einem gut besuchten Hörsaal stattfindet, und auf gute Lesbarkeit der verwendeten Medien. Der Professor dieses Pilotprojektes verwendet in den Zusammenfassungen sowohl die Tafel als auch einen Overheadprojektor und die Projektion von Computersimulationen. Dadurch muss auf wechselnde Beleuchtungsformen eingegangen, die Kamera geschwenkt und gegebenenfalls immer wieder einmal ein Bildausschnitt vergrößert werden. Eine automatische Aufzeichnung würde somit zu einem merklichen Qualitätsverlust führen.

Mit der Verwendung der Aufzeichnung im nächsten Semester könte das Vorlesungselement "Zusammenfassung" bereits als invertiert bezeichnet werden, da es dann nicht nur zur Nach- sondern auch zur kurzen inhaltlichen Vorbereitung dienen kann. Durch die Verwendung der Zusammenfassungen enthalten die Videos die zentralen Inhalte der kompletten Vorlesung, jedoch in extrem kompakter Form.

### 1.3.2  „Partiell-Invertiert": Inverted Classroom zu einem einzelnen komplexen Thema

Die Vermittlung komplexer Zusammenhänge scheitert oft an unterschiedlichen Lerngeschwindigkeiten bei den Studierenden. Zudem gibt es Themen, die auch auf Nachfrage kaum umformuliert sondern meist nur wiederholt werden können, da die Information so kompakt verpackt ist, dass eine Umformulierung die Aussage verfälscht (z.B. Definitionen).

Für ein solches Thema sollte ein hochwertiges Lehrvideo erstellt und multimedial z.B. durch Animationen ergänzt werden, welches die Studierenden im Voraus so oft ansehen sollen bis der Inhalt präsent ist. In der zugehörigen Vorlesung wird der Stoff dann vertieft und das Verständnis durch Anwendungsbeispiele und Abgrenzung erhöht. Die invertierte Einheit ist in dieser Variante kürzer als eine komplette Vorlesung, und im Gegensatz zur vorherigen Version ist auch der aufgezeichnete Stoffumfang stark reduziert. Invertiert wird eine derart multimedial unterstützte Veranstaltung dadurch, dass die Nutzung des bereitgestellten Lehrmaterials als Bedingung für ein Verständnis der in der Präsenzphase bearbeiteten Inhalte zwingend erforderlich ist.

Zielgruppe dieser Variante sind technologisch interessierte Professoren mit hohem Anspruch an die Qualität der Präsentation. Sind an der Hochschule Studiengänge oder Veranstaltungen mit künstlerischem oder redaktionellem Hintergrund vertreten, empfiehlt es sich, zu Beginn einen Kontakt herzustellen, da die Erstellung oft im Rahmen von Projekten oder Studienarbeiten erfolgen kann. Allgemein lohnt es sich, vor der Erstellung neuer multimedialer Inhalte im Internet nach bereits vorhandenen und frei verwendbaren multi-medialen Elementen zu suchen.

Die Umsetzung dieser Form der ICM benötigt mehr Zeit und wird deswegen im laufenden Sommersemester 2012 noch nicht im Rahmen des SKATING Projektes implementiert. Der Professor, auf dessen Anfrage das erste Projekt initiiert wurde, steht jedoch bereits in Kontakt mit möglichen Partnern an der Hochschule.

### 1.3.3 „Zeitweise komplett invertiert": Inverted Classroom in Höherer Mathematik

Die Mathematik wird für viele Studenten gerade im Bereich der Ingenieurwissenschaften zur entscheidenden Hürde Auch auf Seiten der Dozenten kann nach mehreren Jahren Unterricht der Höheren Mathematik der Wunsch nach Veränderung entstehen, da sich der Ablauf der Vorlesung kaum noch verbessern oder erweitern lässt.
Die videobasierte ICM bietet der Vermittlung von Mathematik Unterstützung in zwei zentralen Punkten:
- es ist ein offenes Geheimnis, dass mathematisches Können vor allem durch Anwendung bzw. Übung erlangt wird; daher sollte diesem Aspekt möglichst viel Zeit gewidmet werden.
- zur Wissensvermittlung eignet sich das schrittweise Entwickeln, wie es meist vom Professor an der Tafel durchgeführt wird (Lernen durch 'Abschauen', vgl. Spannagel, 2012).

**Auswahl von Veranstaltung und Thema**

Nach der Vorstellung der Methode durch einen SKATING Mitarbeiter hat ein Professor an der Hochschule Karlsruhe seine *Vorlesung Höhere Mathematik 3* im Sommersemester 2012 explizit der Erprobung des ICM in der Ingenieurmathematik gewidmet. Besondere Beachtung kam dabei der Auswahl des Kurses und des Pilotthemas zu.

Mit etwa 20 motivierten Studierenden aus dem dritten Semester war der fortgeschrittenste der in Frage kommenden Kurse relativ klein. Zudem konnte von den studienerfahreneren Studierenden ein verantwortungsvoller Umgang mit der Methode sowie eine reflektierte Bewertung erwartet werden.

Um die Methode zu erproben hat sich der Professor entschlossen, zunächst nur die Vorlesungen zu einem bestimmten Themengebiet mit vielen Übungsmöglichkeiten zu invertieren. Das Thema *Differentialgleichungssysteme* wird traditionell als eher schwer wahrgenommen, kann aber mit einem gewissen Maß an Vertrautheit und Übung gut gemeistert werden. Klausuraufgaben zu diesem Thema sind meist schematisch lösbar und bilden eine geeignete Basis zum Bestehen des Kurses. Auch die Lage innerhalb der zeitlichen Semesterplanung war geeignet (ca. 3 Wochen nach Semesterbeginn). Ein weiteres (später im Semester behandeltes) Thema (mehrdimensionale Integration) wurde identifiziert. Es kann bei Interesse und positiver Zwischenevaluation ebenfalls invertiert werden. Auch dieses Thema erscheint Studierenden schwieriger als nötig und kann durch Übung gut erfasst werden.

**Aufzeichnung**

Da der Zeitaufwand durch die Auswahl eines einzelnen Themengebietes begrenzt bleibt und die Motivation des Lehrenden hoch war, wurde vereinbart, eine von den regulären Vorlesungen getrennte Aufzeichnung vorzunehmen.

In einem ersten Schritt wurden verschiedene Aufzeichnungsformate mit dem Kandidaten getestet (siehe Abb. 1), um die für ihn bestgeeignete auszuwählen:

- Aufzeichnung einer Vorlesung an der Tafel (vgl. Format von Prof. Spannagel [INT02]).
- Aufzeichnung einer Vorlesung am Overheadprojektor.
- Aufzeichnung der Mitschrift an einem Grafiktablett (vgl. Format von Prof. Loviscach [INT03]).
- Aufzeichnung des Grafiktabletts kombiniert mit einem Video des Dozenten.
- Aufzeichnung der Mitschrift an einem Tablet-Computer kombiniert mit einem Video des Dozenten

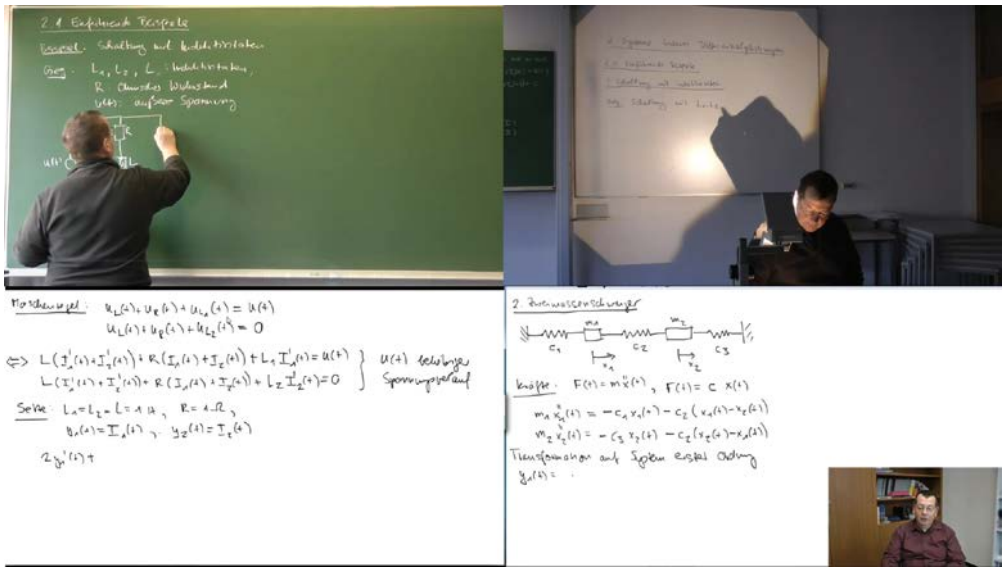

Abb. 2.1:   Beispiele der erprobten Videoformate

Das Format 'Tafel' erschien sehr stilvoll, war jedoch mit dem Problem verbunden, dass für längere Phasen dem Zuschauer der Rücken zugekehrt wird. Bei der Version 'Overheadprojektor' fällt besonders die „künstlerisch" wirkende dunkle Hand ins Auge, der Dozent selbst wird auch ausreichend ausgeleuchtet. Nicht ideal sind jedoch die starken Helligkeitsunterschiede in der Präsentationsfläche und der große ungenutzte Raum an den seitlichen Bildrändern.

Die Entscheidung fiel auf die Aufzeichnung der Mitschrift an einem Grafiktablett in Kombination mit einer Videoaufzeichnung des Professors für die persönliche Ansprache. Dabei wurde Wert auf eine ansprechende Tonqualität gelegt. Der Dozent betonte bei den ersten Versuchen, dass für dieses Format eine eigene Form der Didaktik gefunden werden musste, da das gewohnte Feedback aus dem Auditorium fehlte. Beispielsweise werden eingeschobene Aussprüche wie "ja, sie vermuten richtig, die Antwort ist..." verwendet um das fehlende Publikum einzubeziehen.

Für die Aufzeichnung der Mitschrift wurde zuerst ein Grafiktablett der Marke Bamboo benutzt. Die nach Auskunft des Lehrenden beste Lösung bot nach weiteren Versuchen allerdings ein Tablet-Computer der Marke Lenovo, da durch die direkte Sichtbarkeit des Geschriebenen auf dem Display des Tablet-Computers das Schreiben dem gewohnten Umgang mit Papier entsprach. Für die Aufzeichnung wurde das Programm Camtasia Studio verwendet. Die Videos wurden in HD-Qualität produziert. Verschiedene Schreib- und Zeichenprogramme wurden ausprobiert, wobei auf eine unverfälschte Darstellung der Handschrift geachtet wurde (Freihandzeichnen ohne automatische Konvertierung in Bezier-Kurven). Schließlich dienten die Programme *MyPaint* und *MS OneNote* als ‚Leinwand'. Weitere Kriterien zur Auswahl der Software sind die Möglichkeit zu blättern oder zu scrollen, ein natürlich erscheinender ‚Stift' und die leichte Erreichbarkeit eines Radierers oder weiterer Farben zur Hervorhebung von Inhalten.

Der Lehrende war bereit, die Aufzeichnung komplett alleine durchzuführen, einschließlich dem Aufstellen der Kamera und der Aufnahme des Mitschriebs. Um die Bedienung der Kamera und der Software zu erlernen, wurde eine kurze (30 min) Schulung organisiert. Dieses Format reduzierte nicht nur den Aufwand für die SKATING- Mitarbeiter sondern bietet dem Lehrenden zudem allgemein den Raum, Varianten in eigener Regie auszuprobieren und im weiteren Verlauf unabhängig von der Unterstützung anderer zu werden, falls er es wünscht.

Die Aufzeichnungen wurden danach von einem SKATING-Mitarbeiter bearbeitet. Um ein Video aufzunehmen und zu produzieren wurde insgesamt circa 1 Stunde Arbeitszeit aufgewendet. Der zeitliche Mehraufwand des Dozenten bei eigenständiger Aufnahme lag pro Aufnahmesitzung bei ca. 30 min für das Einrichten des Systems. Zusätzlich wurden ca. 30 min für Organisatorisches aufgewendet. Als herausfordernd erwies sich zudem die Auswahl einer geeignet ruhigen Umgebung, weshalb die Aufnahme letztendlich in den Abendstunden erfolgte. Die Einrichtung eines ‚Studio'-Raumes könnte aus dieser Perspektive lohnend erscheinen.

**Durchführung**

Die erste nach diesem Schema invertierte Veranstaltung wurde bereits durchgeführt. Die Studierenden wurden frühzeitig einbezogen, vorab über die Methode und ihre Ziele unterrichtet und nach ihren Meinungen und Befürchtungen befragt. Die Rückmeldungen waren zumeist positiv, geäußerte Bemerkungen bezogen sich auf die fehlende Möglichkeit Fragen zu stellen und den Anschrieb zu korrigieren, sowie den vermeintlich zusätzlichen Zeitaufwand.

Die Aufzeichnung des drei Vorlesungsstunden umfassenden Stoffes wurde in acht thematische Einheiten von 7-18 min Länge unterteilt und über einen YouTube-Direktlink auf der an der Hochschule verwendeten Lernplattform ILIAS bereitge-stellt. YouTube wurde als Plattform gewählt, da es auf unterschiedlichen, auch mobilen, Endgeräten einwandfrei funktioniert und damit den Studierenden die Möglichkeit anbietet, die Videos fast überall zu schauen. Zweiter Vorteil ist das umfangreiche Analytics-Tool in YouTube, das die Möglichkeit anbietet, Nutzungsstatistiken wie Zugriffe sehr genau zu analysieren.

Da es sich um eine Testphase handelte, wurde auf Wunsch des Lehrenden darauf verzichtet, die Videos komplett zu veröffentlichen. Es wurde daher die Option „nicht gelistet" in YouTube verwendet, wodurch die Videos nur mit dem exakten Link aufzurufen sind und nicht durch Suchmaschinen gefunden werden. Nach einer erfolgreichen Testphase ist geplant, die Videos auch für ein breites Publikum zu veröffentlichen.

Stichworte zu Inhalt und zur Klausurrelevanz der einzelnen Sequenzen sowie ein Forum für die Studierenden zum Anbringen inhaltlicher Fragen wurden im Kursbereich der Lernplattform ILIAS bereitgestellt. Sinnvoll sind außerdem kurze inhaltliche Fragen zu den Videos, die den Lernenden als Leitfragen dienen können und bei der Evaluation als Hinweis auf das erreichte fachliche Verständnis sowie die dem Video gewidmete Aufmerksamkeit zur Verfügung stehen. In ILIAS wurde zudem eine Onlineumfrage zum Nutzungsverhalten und selbsteingeschätzten didaktischen Nutzen der Veranstaltungsform vorgegeben. Es wurden Fragen zum Auftreten von technischen Problemen, zur Bewertung der Videoqualität, zur Videorezeption (u.a. situative Bedingungen, Störungen, Wiederholung), zur Themenpassung und zur wahrgenommenen Nützlichkeit der Videos gestellt.

**Präsenzveranstaltungen**

Die Präsenzveranstaltungen wurden von einem SKATING-Mitarbeiter besucht und anhand eines Beobachtungsbogens das Prozessgeschehen erfasst (z.B. Aufbau und Ablauf der Veranstaltung, Lehrmethoden, Redeanteile, Mitarbeit, Akzeptanz). In der ersten Präsenzveranstaltung erfragte der Lehrende zunächst per Handzeichen die Videonutzung. Die meisten Studierenden hatten sich alle Videos vor der Veranstaltung angesehen (gemäß Frage mit Handzeichen und Nutzungsstatistiken), eine Person hatte technische Probleme bei der Verwendung von ILIAS. Obwohl die im Video erstellten Notizen zum Download zur Verfügung standen, haben Studierende vom Video mitgeschrieben. Da in der zugehörigen Veranstaltung zunächst keine inhaltlichen Fragen von den Studierenden gestellt wurden, erfragte der Professor die relevantesten Punkte aus den Videos, welche korrekt widergegeben werden konnten. Die weitere Präsenzveranstaltung umfasste große Phasen mit betreutem Selbst- und Partnerrechnen sowie eine anschließende Präsentation der Ergebnisse an der Tafel, entweder durch Studierende (auf freiwilliger Basis) oder durch den Professor 'auf Zuruf' vom Auditorium.

**Zielfindung und Evaluation**

Für die Bewertung des Erfolges der Lehrmethode wurde den persönlichen Zielen des Lehrenden ein hoher Stellenwert eingeräumt. In den Vorbesprechungen zur Klärung der Ziele und Ergebniskriterien gab der Lehrende an, dass seine persönlichen Vorstellung einer gelungenen Veranstaltung ein beidseitiges 'Flow'-Erlebnis und eine bei den Studenten geweckte Neugier beinhalte. Eine Vorlesung soll unter dieser Perspektive nicht alle Fragen klären, sondern das Interesse an weiteren Fragestellungen wecken.

Bei der Bewertung des Methodeneinsatzes wird daher insbesondere auch Wert auf den Aspekt der Motivierung der Studierenden gelegt werden. Zusätzlich zu den Onlineumfragen direkt im Anschluss an die Videobetrachtung und den Beobachtungen der Präsenzveranstaltungen sind mindestens zwei weitere Befragungen geplant. Im Anschluss an den ca. drei Vorlesungen umfassenden invertierten Themenkomplex wird im Sinne einer Zwischenevaluation mit einem Fragebogen insbesondere die Zufriedenheit der Studierenden erfasst werden. Bei positiver Resonanz ist, wie oben skizziert, geplant den zweiten Themenblock zu invertieren. Die Bewertung des Erfolges ist damit nicht abgeschlossen, Eine weitere Erhebung unter Einbeziehung unterschiedlicher Ergebniskriterien in der Prüfungszeit und die Analyse der Klausurergebnisse, insbesondere auch im Vergleich zum Vorjahr, soll weitere Hinweise bezüglich des Erfolgs der ICM und dem eventuellen Auftreten von Moderatoreffekten liefern. Weitere Follow-up-Messzeitpunkte können helfen einzuschätzen, inwieweit der gelehrte Stoff nachhaltig angeeignet wurde.

Die Ergebnisse sollen anschließend gemeinsam mit dem beteiligten Lehrenden durchgesprochen werden. Durch die umfassende Informationserhebung soll es ermöglicht werden, den Einsatz und den Erfolg der ICM in diesem Anwendungskontext angemessen zu beurteilen. Auf Basis dieser Beurteilung ist dann das weitere Vorgehen, sind die nächsten Schritte zu planen.

Eine positive Beurteilung kann die Ausgangsbasis für weitere wissenschaftlich begleitete Pilotprojekte mit der ICM sein. In diesen können die Erfahrungen und identifizierten Verbesserungspotenziale des ersten Projekts umgesetzt werden.

## 1.4 Ausblick

Bei positiver Evaluation der Pilotprojekte, bei vertretbarem Aufwand und Interesse seitens der Dozierenden und Studierenden soll weiter über das Konzept des Inverted Classroom informiert werden. Das Projekt SKATING kann dadurch als Multiplikator dienen und zugleich bei der Anpassung an Lehrende und Lehrveranstaltungen sowie bei der Durchführung und Evaluation konkrete Unterstützung bieten. Über die offene Kommunikation der Ergebnisse und durch Schaffung der Rahmenbedingungen könnte die ICM so fester Bestandteil des Methodenrepertoires der Lehrenden an der Hochschule Karlsruhe werden.

Wenn ausreichend Erfahrungen mit unterschiedlichen Veranstaltungen gesammelt wurden, können dadurch in vergleichenden Studien fundierte und auch für andere Hochschulen interessante Schlüsse gezogen werden.

## 1.5 Schlussfolgerungen

Das gegenwärtige Lehrsystem scheint so schlecht doch gar nicht zu funktionieren. Immerhin reproduziert sich die akademische Elite auf diese Weise. Alle Professorinnen und Professoren wurden auf diese Weise trainiert. Wieso sollten wir das System der Lehre dann aber überhaupt reformieren?

Weil es sozialdarwinistisch („Survival of the Fittest") und zynisch erscheinen mag. Weil es einen bestimmten Studierendentypus voraussetzt oder mittels eines personalaufwändigen Übungsbetriebes erzwingt, der auch bei größter Überforderung noch frustrationstolerant seine Problemlösungskompetenz in durchwachten Nächten trainiert. Weil es allem Anschein nach eine große Zahl von Studienabbrechern produziert. Weil gerade die Hochschulen für Angewandte Wissenschaften den ge-sellschaftlichen Auftrag haben, traditionell bildungsfernere Schichten für ein Studium nicht nur zu motivieren, sondern dieses dann auch studierbar zu machen.

Es gibt also genügend Gründe, sich um einen alternativen Umgang mit der Vorlesung zu bemühen ohne sie pauschal zu diskreditieren. Gerade die Methode des IC scheint geeignet, strukturelle Schwächen der Vorlesung aufzuheben oder zu kompensieren ohne dabei auf viele ihrer Vorteile verzichten zu müssen. Durch die Kombination von E-Learning Elementen mit einer individuell unterstützenden Präsenzveranstaltung hat diese Methode mit Vergangenheit das Potential in sich, auch zu einer wichtigen Lehrmethode der Zukunft zu werden.

Unter IC lassen sich in der konkreten Umsetzung eine Reihe unterschiedlicher Varianten und Ausgestaltungsmöglichkeiten fassen, von denen einige vorgestellt wurden. Wesentliche Merkmale und durch den Einsatz der Methode erwartete positive Effekte sind jedoch den meisten Modellen gemein.

Im spezifischen Anwendungskontext sollte die Methode sehr genau auf die individuellen Ziele, Bedürfnisse und Möglichkeiten von Lehrenden und Lernenden zugeschnitten werden um den größtmöglichen Erfolg zu erzielen. Hier sollte weder vor individuellen Lösungen noch vor der Verwendung bereits vorhandener Materialien zurückgeschreckt werden. Es ist von Vorteil, wenn den Lehrenden Unterstützung geboten werden kann, welche sich bestenfalls von der Entwicklung einer passenden Methode über Hilfen bei der technischen Umsetzung bis zur begleitenden Beobachtung und Evaluation erstreckt, damit sich der Lehrende auf die so wesentliche Ausgestaltung der Präsenzphase konzentrieren kann.

Wichtig ist hierbei, jeweils die Rahmenbedingungen und Effekte sehr genau zu erfassen und zu dokumentieren, damit die Ergebnisse in zukünftigen Metaanalysen aussagekräftig verwendet werden können.

Professoren, die besonders gut mit dem Medium der klassischen Vorlesung arbeiten und Studierende mitreißen können, könnten Ihre Vorlesungsaufzeichnungen zur Verfügung stellen. Diese könnten ohne Prestigeverlust als zusätzliche Ressource für andere Veranstaltungen dienen, wenn eine Kultur des Teilens unterschiedlichster Lehr- und Lernmaterialien entsteht. Gute Beispiele sprechen sich herum und steigen in den Trefferlisten der Suchmaschinen. Erfolgreiches Lernen (und Lehren!) kann Spaß machen, Neugier wecken und die entsprechenden Lehrenden haben das Potential zu „Role Models" für exzellente Vermittlung zu werden.

Abschließen soll ein Kommentar eines Studierenden, denn dieser fasst die möglichen Vorteile, das Potential der ICM für die Studierenden und den Charme dieser Methode treffend zusammen: „Man kann jetzt in der Vorlesung mitreden"!

# Literatur